\documentclass[aps,prl,twocolumn,superscriptaddress,showpacs]{revtex4}
\usepackage{amsfonts,amssymb,amsmath}
\usepackage{color}
\usepackage{graphicx}
\usepackage{dcolumn}
\usepackage{bm}

\newcommand{\ket}[1]{|#1 \rangle}
\newcommand{\Tr}{\mathop{\mathrm{Tr}}}

\begin{document}

\title{Quantum Monte Carlo simulations of fidelity at magnetic quantum phase transitions}

\author{David Schwandt}
\author{Fabien Alet}
\author{Sylvain Capponi}
\affiliation{Laboratoire de Physique Th\'eorique, Universit\'e de Toulouse, UPS, (IRSAMC), F-31062 Toulouse, France}
\affiliation{CNRS, LPT (IRSAMC), F-31062 Toulouse, France}

\date{\today}

\begin{abstract}
 When a system undergoes a quantum phase transition, the ground-state
 wave-function shows a change of nature, which can be monitored using the {\it fidelity} concept. We introduce two Quantum Monte
 Carlo schemes that allow the computation of fidelity and its
 susceptibility for large interacting many-body systems. These
 methods are illustrated on a two-dimensional Heisenberg model, where
 fidelity estimators show marked behaviours at two successive quantum phase
 transitions. We also develop a scaling theory which relates the divergence of
 the fidelity susceptibility to the critical exponent of the correlation
 length. A good agreement is found with the numerical results.
\end{abstract}


\maketitle

What happens to the ground-state (GS) wave-function when a physical
system goes across a quantum phase transition (QPT)? Rooted in quantum
information theory, the fidelity approach~\cite{Zanardi,FidelityReview}
provides an interesting global answer in terms of the overlap between
GS of the system at two different values of the driving
parameter. The basic idea, which precursor may be found in Anderson's
orthogonality catastrophe~\cite{Anderson}, is that close quantum states
become more orthogonal close to a phase transition. The resulting fidelity drop can then provide a useful probe to detect the
QPT. This is particularly interesting as the fidelity is a global, model-independent quantity that incorporates all
the information contained in the GS wave-functions. This is opposite
to other usual approaches to phase transitions, which often need an
input such as the knowledge of a specific order parameter.

Consider the Hamiltonian
\begin{equation}
H = H_0 + \lambda H_\lambda
\label{eq:Ham}
\end{equation}
with $H_\lambda$ acting as a perturbation to $H_0$.  The fidelity $F(\lambda_1,\lambda_2)$ is defined as the
modulus of the overlap between GS of $H$ at two different
values of $\lambda$:
\begin{equation*}
F(\lambda_1,\lambda_2)=|\langle  \psi_0^{\lambda_1} | \psi_0^{\lambda_2}
\rangle |.
\end{equation*} 
Suppose that the system undergoes a QPT
for a value $\lambda_c$ of the driving parameter.
We expect the fidelity to have a singular behaviour when either $\lambda_1$
or $\lambda_2$ are close to $\lambda_c$, especially if the
difference $\delta \lambda=\lambda_2-\lambda_1$ is small~\cite{Zanardi}. When
$\delta \lambda\rightarrow 0$, the fidelity is dominated by its leading
term, the fidelity susceptibility $\chi_F$~\cite{chiint}, with $F\simeq 1-\frac{\delta \lambda^2}{2}\chi_F$.

When $H$ describes a many-body system, computations of $F$ or
$\chi_F$ are complicated. Besides a few analytical results on specific
models~\cite{FidelityReview}, the main effort has been put in their
numerical evaluation. Exact diagonalization (ED) and
tensor-network (TN) methods~\cite{Tensor} -including density matrix
renormalization group (DMRG)~\cite{DMRG}- have been the most widely
used techniques in that respect, even though they suffer from several
caveats. The ED method needs the full computation
of the GS wave-function and is therefore limited to small
systems. TN methods provide a variational ansatz for the
GS wave-functions, allowing a straightforward computation of
overlaps. This ansatz turns out to be excellent for one-dimensional systems,
where DMRG~\cite{DMRG} in particular has proved its full strength. Recently, several TN based works studied two-dimensional
(2$d$) systems and their fidelity properties~\cite{Tensor2d}. However,
these methods remain variational and may fail in correctly capturing
the GS properties of complex many-body Hamiltonians in $d>1$, especially
close to a QPT.

In this Letter, we present two different Quantum Monte Carlo (QMC)
schemes which allow an {\it exact} (albeit stochastic) computation of
the fidelity $F$ and its susceptibility $\chi_F$. We apply these
methods to the antiferromagnetic (AF) Heisenberg spin model on a 2$d$
lattice. Varying one exchange coupling in the spin model
causes two successive phase transitions, both of which are found to be
captured by the fidelity and its susceptibility. In passing, we
derive a scaling theory for the divergence of $\chi_F$ at a
second-order QPT. The two schemes benefit from the power of QMC methods,
which allow to treat very large systems \emph{in any dimension}. The first
scheme, which calculates $F$, is applicable to all AF systems admitting a
singlet GS. The second scheme for $\chi_F$ is even more general and can be
applied to several many-body problems. This
opens a novel path for the fidelity approach to QPT. These techniques are only efficient when the underlying QMC
method is, {\it i.e.} when there is no sign problem.

\begin{figure}
\includegraphics[height=4cm]{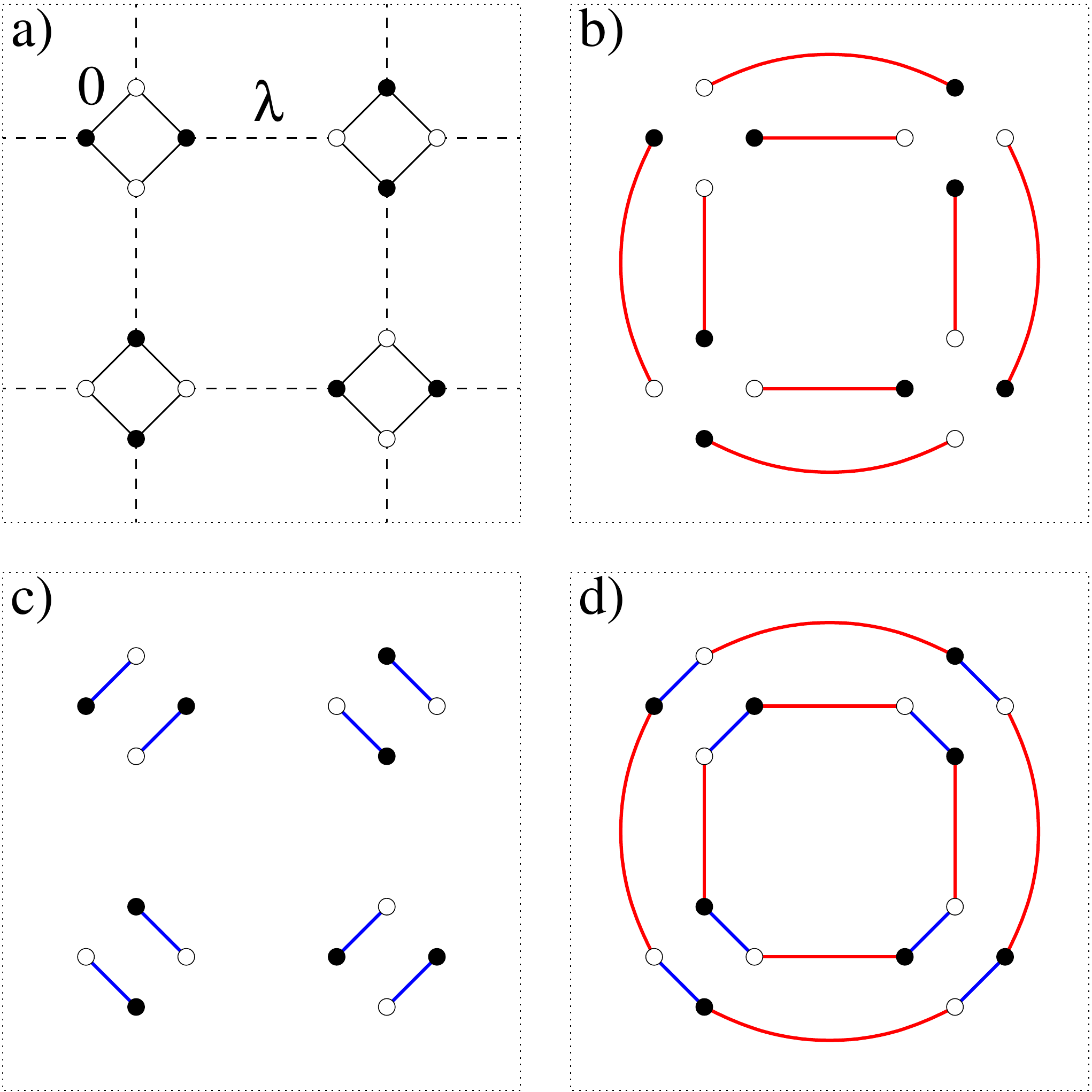}
\caption{(color online) (a) CaVO lattice with nearest neighbor bonds
 of type $\lambda$ (dashed lines) and $0$ (solid lines). (b)-(c) Typical
 VB states on the CaVO lattice. (d) Overlap graph of previous VB states
 forming two loops.}
\label{fig:CaVO}
\end{figure}

{\it Model --- }
The schemes are illustrated on the spin-$1/2$ Heisenberg Hamiltonian
\begin{equation}
\label{eq:H}
H=H_0+\lambda H_\lambda=\sum_{\langle ij \rangle_0} {\bf S}_i\cdot {\bf S}_j
+ \lambda \sum_{\langle ij \rangle_\lambda} {\bf S}_i\cdot {\bf S}_j
\end{equation}
on the CaVO lattice~\cite{Cavo}, a 1/5th depleted square lattice (see
Fig.~\ref{fig:CaVO}). The first sum runs over
nearest-neighbor spins on $0$-bonds (solid lines in
Fig.~\ref{fig:CaVO}) while the second is over $\lambda$-bonds (dashed
lines). This lattice structure can be found in the AF compound
CaV$_4$O$_9$ (hence the lattice name), even though the interactions are more complex in the real
compound~\cite{CavoExp}. Varying the coupling $\lambda$
allows the occurrence of two QPT separating an
intermediate N\'eel-ordered AF phase from respectively a low-$\lambda$
plaquette and a high-$\lambda$ dimer phase~\cite{Cavo,QMCCavo}. $H$ conserves the total spin of the
system and in particular, AF interactions $\lambda>0$ lead to a
singlet GS.

{\it Fidelity measurements --- } Measuring the fidelity seems at first
glance easy within a standard Projection QMC
scheme~\cite{QMCProj}. Decomposing the ground-state
$\ket{\psi_0^\lambda}=\sum_i a_i \ket{\varphi^\lambda_i}$ in the
simulation basis $\{\ket{\varphi}\}$, one generates representatives
$\ket{\varphi^\lambda_i}$ ({\it i.e.} in proportion of $|a_i|$) of the
GS via the projection scheme for two different values of
$\lambda$. The problem comes from the fact that the QMC
estimate of the fidelity $\langle \varphi^{\lambda_1}_i |
\varphi^{\lambda_2}_i \rangle$ will vanish most of the time in the
commonly used orthogonal basis, leading to a serious statistical
problem. However, when the GS is a singlet, it can be decomposed
in the Valence Bond (VB) basis, which has a crucial 
non-orthogonality property. Indeed, any two VB states always have a
non-zero fidelity $F=| \langle \varphi_1 | \varphi_2 \rangle
|=2^{N_{\ell}-N/2}$ where $N_{\ell}$ is the number of loops obtained by
superimposing the two VB states (see Fig.~\ref{fig:CaVO}), and $N$ the
total number of spins. This property solves the statistical
problem and allows an efficient computation of the fidelity.

More specifically, we work with a VB projector loop algorithm recently
proposed by Sandvik and Evertz~\cite{SandvikEvertz}. To avoid the sign problem, we simulate
non-frustrated AF on bipartite lattices, which leads to real positive values of coefficients
$a_i\geq 0$ and of VB overlaps $\langle \varphi_1 | \varphi_2
\rangle>0$. In the VB loop algorithm~\cite{SandvikEvertz}, two VB
representatives $\ket{\varphi_L}$ and $\ket{\varphi_R}$ of the
ground-state are generated by propagating two initial VB
states. Simulating at the same time {\it two} different physical
systems with couplings $\lambda_1$ and $\lambda_2$ allows a QMC estimator of the square of the fidelity:
\begin{equation*}
F^2 (\lambda_1,\lambda_2)= \frac{\langle \varphi_L^{\lambda_1} | \varphi_R^{\lambda_2}
 \rangle \langle \varphi_R^{\lambda_1} | \varphi_L^{\lambda_2} \rangle}{\langle \varphi_L^{\lambda_1} | \varphi_R^{\lambda_1}
 \rangle \langle \varphi_L^{\lambda_2} | \varphi_R^{\lambda_2} \rangle}.
\end{equation*}

$F(\lambda_1,\lambda_2)$ can be computed for any value of $\lambda_1$
and $\lambda_2$ for all models that can be simulated with VB QMC
methods. 

In the following, we illustrate this method for the Heisenberg model on the
CaVO lattice (Eq.~\ref{eq:H}). The unit cell contains 4 spins, and we
simulated square samples with $L\times L$ unit cells (total number of spins
$N=4L^2$) up to $L=16$, using periodic boundary conditions. For such large
systems, the fidelity essentially vanishes for all $\lambda_1 \neq
\lambda_2$. As suggested in Ref.~\cite{FPS}, we compute the fidelity per
site $f(\lambda_1, \lambda_2) =  F(\lambda_1,\lambda_2)^{1/N}$, which is well-behaved as $N\rightarrow \infty$. 

\begin{figure}
\includegraphics[width=\columnwidth]{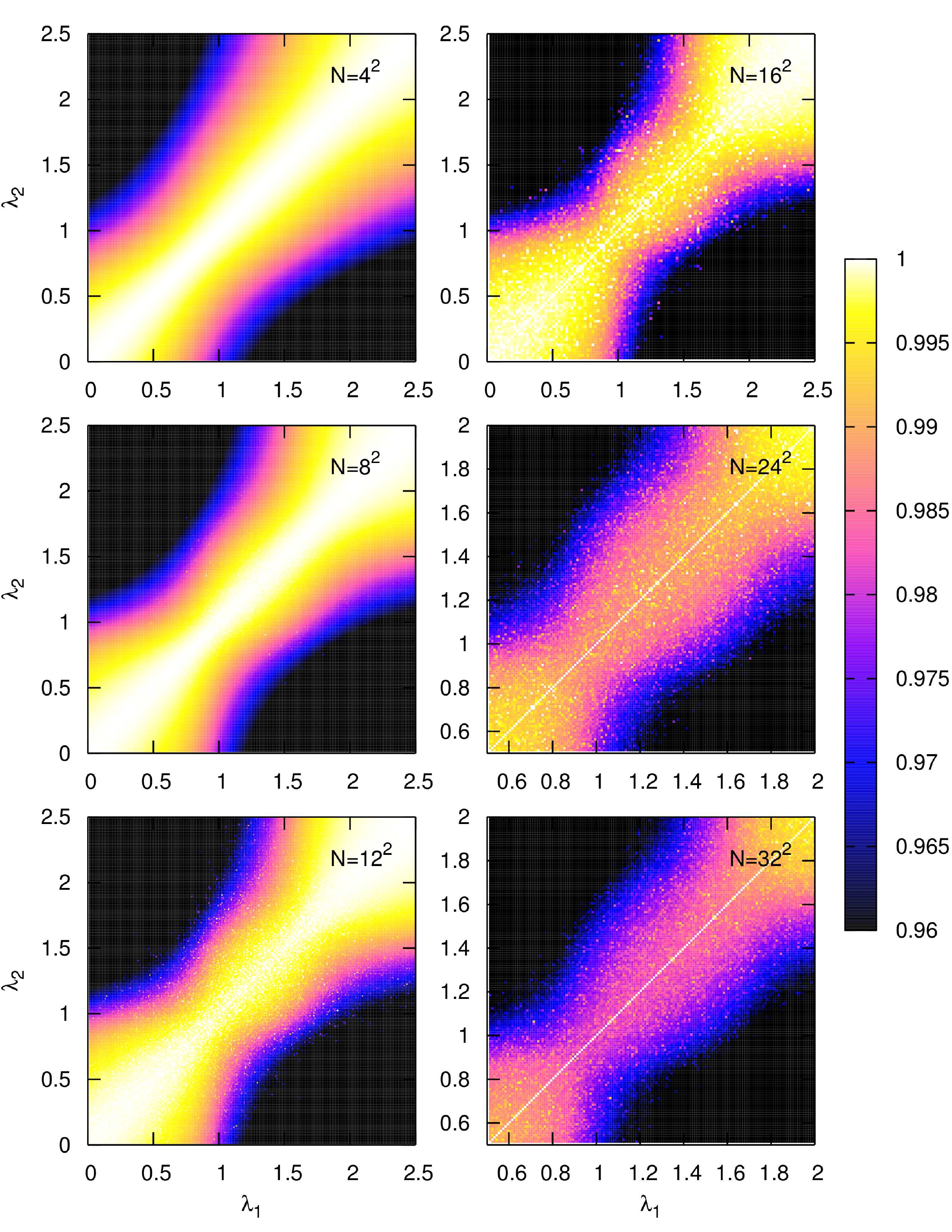}
\caption{(color online) Fidelity per site $f$ as function of
 $\lambda_1$ and
 $\lambda_2$ for different system sizes. Expansion power $n$ of $H$ in the VB QMC method is $n/N=20$ for $N\leq 12^2$,
 $10$ for $N=16^2$ and  $4$ for $N\geq 24^2$. $\lambda$ range is
 $[0,2.5]$ for $N\leq 16^2$ and $[0.5,2]$ for $N\geq 24^2$. Resolution
 $\Delta\lambda$ for the plots is $0.02$ for $N=16^2$, $0.01$ otherwise.
 }
\label{fig:pm3d}
\end{figure}

Our data for the fidelity per site are presented on Fig.~\ref{fig:pm3d}. Around 
the diagonal where $f(\lambda,\lambda)=1$, we notice the appearance 
of two pinch points, roughly around $\lambda_{c}^1\in[0.8,1.1]$ and
$\lambda_{c}^2\in[1.5,1.8]$. It has been argued that these
features are characteristic of continuous QPT~\cite{FPS} and our results are
in agreement with the two well-known second-order QPT in this
model. Far enough away from these two critical regions we notice that there
is no significant change of $f$ when
$N\rightarrow\infty$. Within the critical regions $f$
drops faster with system size. Given our statistical errors (up to $2\%$ for the chosen range of
$\lambda$), we cannot however provide more precise ranges for the critical
points. To locate more accurately the QPT, we now turn to the leading correction of fidelity around the diagonal $\lambda_1=\lambda_2$.

{\it Fidelity susceptibility --- }
For $\delta \lambda\rightarrow 0$, we consider the fidelity
susceptibility $\chi_F$ which can be expressed~\cite{chiint} as the
imaginary-time integral $\chi_F=\int_0^\infty \tau \left[ \langle H_\lambda (0)
 H_\lambda(\tau)\rangle -\langle H_\lambda (0) \rangle^2\right]
d\tau.$ This definition offers a natural
extension to finite temperature $T=1/\beta$:
\begin{equation}
\chi_F(\beta)=\int_0^{\beta/2} \tau \left[ \langle H_\lambda (0)
 H_\lambda(\tau)\rangle -\langle H_\lambda (0) \rangle^2\right]d\tau.
\label{eq:chi}
\end{equation}
and $\chi_F=\lim_{\beta\rightarrow \infty}\chi_F(\beta)$. This definition of
$\chi_F(\beta)$ differs from the Bures metric $ds^2$ usually defined for
mixed states~\cite{Bures}, even though both have the same $T=0$
limit. However, one can prove~\cite{prep} that $ds^2/2\leq \chi_F(\beta)
\leq ds^2$, showing that both quantities scale in the same way. 

An advantage of Eq.~(\ref{eq:chi}) is that $\chi_F(\beta)$ can be
computed within a QMC stochastic series
expansion (SSE) formalism~\cite{Sandvik92,Sandvik99}. Note the importance of
taking $\beta/2$ as the upper limit of the integral as the
$\beta$-periodicity of the path integral would lead otherwise to incorrect results. In the SSE formalism the partition function is expanded in
powers of $\beta$, $Z=\sum_{n=0}^{\infty} \frac{(-\beta)^n}{n!}\Tr
H^n$. Starting from the expression of time-displaced correlations
functions in SSE (Eq. 3.15 of Ref.~\onlinecite{Sandvik92}), $\chi_F(\beta)$ is estimated as
\begin{equation}
\chi_F(\beta)=\frac{1}{\lambda^2}\left[ \sum_{m=0}^{n-2} A(m,n)\langle N_\lambda(m) \rangle -
 \langle N_\lambda \rangle^2/8 \right]
\label{eq:chisse}
\end{equation}
where $N_\lambda(m)$ is the number of times two elements of
$H_\lambda$ appear separated by $m$ positions in the SSE sequence~\cite{Sandvik92} and
$N_\lambda$ the total number of appearance of elements of
$H_\lambda$. 
The amplitude $A(m,n)=\frac{(n-1)!}{(n-m-2)!m!}\int_0^{1/2} d\tau
\tau^{m+1}(1-\tau)^{n-m-2}$ can be calculated  for all $(m,n)$ prior to
simulations by numerical integration or analytically for large $n$~\cite{prep}. We emphasize that
this formalism allows to compute $\chi_F(\beta)$ for any model which
can be simulated with SSE.

The computation of $\chi_F(\beta)$ can turn costly for
large systems at low $T$. We reached $L\leq 16$ and used $\beta=10 L$ for the CaVO lattice, and limited simulations to the
relevant $\lambda$ range for the largest $L$ and $\beta$. 

\begin{figure}
\includegraphics[height=7cm]{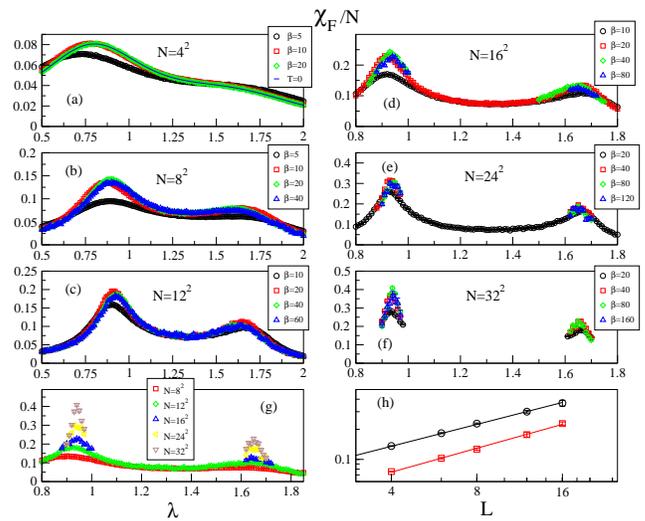}
\caption{(color online) (a)-(f) Fidelity susceptibility per site $\chi_F/N$ versus
 $\lambda$ for different system sizes $N$ and inverse temperature $\beta$. (g)
 $\chi_F/N$ versus $\lambda$ for the largest $\beta$ for different $N$. (h)
 Scaling of the two peaks of  $\chi_F/N$ versus linear size $L$ (log-log scale). Lines
 denote power-law fits.}
\label{fig:chi}
\end{figure}

Fig.~\ref{fig:chi}(a)-(f) display the susceptibility fidelity per
site $\chi_F/N$, showing the apparition of two peaks as a function of $\lambda$. While the $N=4^2$ and $8^2$
samples show rather broad feature (especially for the second peak),
the peaks are clearly emergent as system size is increased and
temperature lowered. From the position of the two peaks for the lowest
$T$ and largest size, one obtains estimates $\lambda_c^1=0.94(1)$ and
$\lambda_c^2=1.65(1)$ for the two quantum critical points, in full
agreement with QMC computations of order parameter and spin
gap~\cite{QMCCavo}. Note that the positions of the maxima of $\chi_F$
at finite $T$ (see Fig.~\ref{fig:chi}(e)-(f)) also allow to
determine faithfully $\lambda_c^1$ and $\lambda_c^2$, even though a
small shift is observed if $\beta$ is too low. We therefore find that the fidelity susceptibility behaves as a
good global indicator of QPTs in a 2$d$ quantum system.

Away from criticality, $\chi_F$ is extensive in all phases - see
Fig.~\ref{fig:chi}(g). Scaling of the peaks in Fig.~\ref{fig:chi}(h) reveals a power-law divergence at criticality
$\chi_F(\lambda_c) / N \sim L^{\omega}$, with $\omega=0.73(3)$ for the first
QPT and $\omega=0.79(6)$ for the second (error bars originate from the
statistical error bar in the QMC data). This agrees with the
behaviour of $f$ which develops pinch points at
criticality but essentially does not change in non-critical regions as
$N\rightarrow\infty$. The observed symmetry of the peaks around their
divergence explains the hourglass shape of the pinch points in $f$.
We also understand why the second pinch point is
harder to see in Fig.~\ref{fig:pm3d}: indeed $\chi_F$ is much smaller close
to $\lambda_c^2$ than to $\lambda_c^1$ (almost a factor of $2$ in all cases).

{\it Scaling theory --- } We now account for the divergence at
the QPT by formulating a finite-size scaling (FSS) theory. FSS theories for $\chi_F$ have been proposed earlier but the
relation to standard critical exponents at second-order QPT has been
missed. In the most elaborate work, Campos Venuti and Zanardi~\cite{FSS}
discuss the divergence of $\chi_F$ as a function of the scaling dimension
$[H_\lambda]$ of the part of the Hamiltonian that drives the transition, {\it
 i.e.} how this operator scales at $\lambda_c$: $H_\lambda \sim L^{-[H_\lambda]}$. Here we
explicitly calculate this scaling dimension. From the definition of the
correlation length critical exponent $\xi \sim (\lambda-\lambda_c)^{-\nu}$,
we have $[\lambda]=1/\nu$. Noting that $[H]=z$ where $z$ is the
dynamical critical exponent, we deduce $[H_\lambda ]=z-1/\nu$ from
Eq.~(\ref{eq:Ham}). Finally, we conclude from Eq.~(\ref{eq:chi}) that
$[\chi_F]=2 [\tau]+2  [H_\lambda ]=-2/\nu$ (see also Ref.~\cite{FSS}) and
therefore $[\chi_F/N]=-2/\nu+d$, leading to the prediction:
\begin{equation}
\chi_F (\lambda_c)/N \sim L^{2/\nu-d}.
\label{eq:scaling}
\end{equation}
This relation should hold for all second-order QPT and explains the
superextensive behaviour generally observed for $\chi_F$ in terms of the
usual critical exponents. From the value $\nu \simeq 0.7112$
 of the universality class of the 3d $O(3)$ model~\cite{Campostrini} to
 which belong both QPT studied here, we expect from this analysis
 $\omega\simeq 0.812$, in agreement with our numerical estimates.

{\it Discussions and conclusion --- } In conclusion, we presented
two QMC schemes that are able to calculate with high accuracy the
fidelity and its susceptibility  for quantum interacting
systems, in any dimension. This allows to pin down the behaviour of
fidelity at QPT, using one of the most
sophisticated numerical techniques for the many-body problem. Taking
the example of the Heisenberg model on the CaVO lattice, we find that
both $F$ and $\chi_F$ are able to locate the two quantum critical
points present in this system. The fidelity susceptibility acts as a
more precise indicator as criticality manifests itself as a marked
peak in $\chi_F$.  However, there is in principle more information
contained in $F$. This could be useful to detect transitions that
$\chi_F$ does not capture~\cite{capture}, as well as in the context
of quantum quenches~\cite{quench,quench2}. 

The divergence of $\chi_F$ at criticality is accounted for by the scaling
theory that we have presented, where the connection to the correlation
length exponent of the universality class of the QPT is made. 
We also showed that the generalization of $\chi_F$ to finite temperature
(Eq.~\ref{eq:chi}) allows to detect criticality for moderate values of
$T$. This is of practical interest as simulations can be performed at a smaller
computational cost. 

The method proposed for measuring $\chi_F$ works for any
model which can be simulated within the generic SSE
scheme~\cite{Sandvik92,Sandvik99}, opening the door to the study of fidelity
in many different physical systems. We expect that our scheme can be
extended to measure the Loschmidt echo, another witness of quantum
criticality~\cite{Loschmidt1}, which can be measured experimentally in this context~\cite{Loschmidt2}.

{\it Note added --- } The scaling relation Eq.~(\ref{eq:scaling}) has been
independently derived in recent preprints~\cite{quench2}.

{\it Acknowledgments --- } We thank F. Albuquerque, A. L\"auchli, O. Motrunich, G. Roux and C. Sire for very useful discussions. Calculations were
performed using the ALPS libraries~\cite{ALPS}. We thank GENCI and CALMIP for allocation of CPU time. This work is supported by the French
ANR program ANR-08-JCJC-0056-01.

\end{document}